# Quantification of locked mode instability triggered by a change in confinement

M. Peterka[1,2], J. Seidl[1], T. Markovic[1], A. Loarte[3], N.C. Logan[4], J.-K. Park[5], P. Cahyna[1], J. Havlicek[1], M. Imrisek[1], L. Kripner[1,2], R. Panek[1], M. Sos[1], P. Bilkova[1], K. Bogar[1], P. Bohm[1], A. Casolari[1], Y. Gribov[3], O. Grover[1], P. Hacek[1], M. Hron[1], K. Kovarik[1], M. Tomes[1,2], D. Tskhakaya[1], J. Varju[1], P. Vondracek[1], V. Weinzettl[1], and the COMPASS Team[6]

[1] Institute of Plasma Physics of the CAS, Prague, Czech Republic
[2] Faculty of Mathematics and Physics, Charles University, Prague, Czech Republic
[3] ITER Organization, St Paul-lez-Durance, France
[4] Columbia University, New York, USA
[5] Princeton Plasma Physics Laboratory, Princeton, USA
[6] See Hron et al 2022 (https://doi.org/10.1088/1741-4326/ac301f) for the COMPASS Team.

E-mail: peterka@ipp.cas.cz

**Abstract**

This work presents the first analysis of the disruptive locked mode (LM) triggered by the dynamics of a confinement change. It shows that, under certain conditions, the LM threshold during the transient is significantly lower than expected from steady states. We investigate the sensitivity to a controlled $n = 1$ error field (EF) activated prior to the L-H transition in the COMPASS tokamak, at $q_{95} \sim 3$, $\beta_N \sim 1$, and using EF coils on the high-field side of the vessel. A threshold for EF penetration subsequent to the L-H transition is identified, which shows no significant trend with density or applied torque, and is an apparent consequence of the reduced intrinsic rotation of the 2/1 mode during this transient phase. This finding challenges the assumption made in theoretical and empirical works that natural mode rotation can be predicted by global plasma parameters and urges against using any parametric EF penetration scaling derived from steady-state experiments to define the error field correction strategy in the entire discharge. Furthermore, even at EFs below the identified penetration threshold, disruptive locking of sawtooth-seeded 2/1 tearing modes is observed after about 30% of L-H transitions without external torque.

## 1. Introduction

The inevitable presence of an error field (EF), a stationary small-amplitude (as low as $\delta B / B \sim 10^{-4}$ [1]) 3D perturbation $\delta B$ of the tokamak magnetic field $B$, arises from the imperfections of the magnetic coils (manufacturing and positioning) and other sources of toroidal asymmetry. A resonant component of the perturbation $\delta B$ can drive magnetic reconnection, producing a tearing mode (TM) at the corresponding rational $q(r) = m/n$ magnetic surface, with the safety factor radial profile $q(r)$, and the poloidal and toroidal mode numbers $m, n$. Fortunately, a stationary $\delta B$ is typically rotationally suppressed by the tokamak plasma until its amplitude exceeds a certain threshold, above which the TM becomes locked to $\delta B$ and full reconnection occurs in an event termed EF penetration [2]. The resulting locked mode (LM) often leads to a disruption - especially in a low-$q_{95}$ plasma with an $m/n = 2/1$ LM [3]. Predicting the threshold for EF penetration in ITER and other future tokamaks is thus of significant importance and has been the subject of extensive research. This work aims to extend it to regimes previously not covered by investigating the EF penetration threshold during the L-H transition.

In many theoretical models, the naturally occurring rotation frequency of a tearing mode in a plasma unperturbed by the EF is the key parameter for its stabilization against its growth by the penetration of a non-rotating EF. In a simplified picture, the electromagnetic torque is balanced by the torque sources opposing any change to the natural mode frequency, such as plasma perpendicular viscosity, which is manifested by the positive proportionality of the penetration threshold to the natural mode frequency [2, 4, 5]. The same conclusion holds for the more elaborate models [6, 7], where, unlike in the earlier studies, the natural mode frequency is no longer considered an unknown free parameter. Instead, the natural mode frequency in Ohmic plasma is assumed to scale like the diamagnetic frequency, with dimensionless parameters $\rho^*$ and $\beta$. This assumption is used to express the final predictions solely in either dimensionless or engineering terms [7].

On the experimental side, empirical multi-machine scaling of the $n = 1$ EF penetration threshold with global discharge parameters has recently been updated, and the $n = 2$ threshold has been formulated, in [8]. In this work, the resonant EF coupling amplitude is quantified by the GPEC overlap field metric, incorporating the plasma response through the ideal MHD approximation [9, 10]. These scalings with global plasma parameters, consistent with the theoretical works, implicitly represent rotation through its correlation with plasma pressure in co-current neutral beam injection (NBI) heated plasma and through an unknown function of all scaling parameters in Ohmic plasma [8].

These assumptions about mode rotation are inherently challenged in regimes beyond those covered by the experimental data set used for scaling. Strong sensitivity of the $n = 1$ EF penetration threshold to the rotation deviating from the assumed trend has been demonstrated through natural changes of rotation in Ohmic plasma in J-TEXT [11], through rotation braking by an additional $n = 3$ EF in high-$\beta$ plasma in NSTX [12], and by using a mix of NBI with different toroidal angles of co-injection in H-mode in DIII-D [13].

Almost all existing studies involving controlled EF have investigated its effects in steady-state conditions, such as when the discharge stabilizes following a change in power and torque input. An instance of introducing EF into a non-stationary phase of the discharge is found in the application of an $n = 1$ EF to an Ohmic ELM-free H-mode in COMPASS-D [14]. Here, EF penetration was observed at an order of magnitude higher density than in Ohmic L-mode under otherwise identical conditions. This phenomenon was attributed to the observed spontaneous drop of the natural 2/1 mode frequency from around 10 kHz to approximately 3 kHz within the first 10 ms of H-mode, and subsequently to less than 1 kHz, regardless of the EF presence. This observation was supported by B3+ ion toroidal rotation measurements. No upper limit in density for EF penetration was identified when the EF was applied at various timings during the natural ramp of density (and $\beta_N$) in the ELM-free H-mode. However, it's worth noting that the high-density points may have been influenced by resonant field amplification at high $\beta_N$.

Recent experiments in the COMPASS tokamak have revealed a remarkable consequence of applying the EF prior to the NBI-assisted L-H transition at $q_{95} \sim 3$. Preliminary results [15] reported the prevention of H-mode access in this scenario due to disruptions induced by an $n = 1$ EF created by a set of 3D coils on the high-field side (HFS). When compensating the dominant 2/1 resonant component of this HFS EF with an additional low-field side (LFS) EF, the disruption rate was only reduced to about 50%, which was ascribed to the non-resonant component of the EF correction scheme [15, 16]. However, the residual 2/1 resonant component remaining after an imperfect correction could also explain the observed disruptions, since the upper limit of plasma sensitivity to EF penetration in non-stationary conditions has, to our knowledge, never been quantified. The present work therefore aims to characterize this sensitivity, considering that the reported effects could have significant implications for EF correction in future tokamaks where a high disruption rate is unacceptable.

It should be noted that the resonant components of a HFS EF couple to the plasma differently than the typical LFS-originated EFs used in [8]. However, no significant difference in EF penetration threshold was observed in Ohmic plasma at $q_{95} \sim 3$ in COMPASS, which will be addressed in a forthcoming paper.

This work reports the first systematic analysis of the disruptive locked mode instability triggered by the dynamics of a transient change in the plasma regime. In particular, the L-H transition in presence of $n = 1$ HFS EF is under study, where the EF current is held constant and the resonant coupling is quantified by the GPEC overlap metric of [8]. The NBI-assisted as well as the Ohmic L-H transitions are included in the data set of the LM occurring just after the L-H transition, with $\beta_N \sim 1$ to exclude resonant field amplification and $q_{95} \sim 3$ to be relevant with the ITER baseline scenario as described in [17]. This differs from many EF studies that have been performed at higher $q_{95}$ so that the LM would typically not be disruptive (e.g., [18]).

This paper is organized as follows. In Section 2, the experimental scenario and diagnostic methods are described in more detail. Section 3 demonstrates that the HFS EF behaves in accordance with the standard resonant coupling experience with respect to the critical density for EF penetration in Ohmic plasma. This serves as a reference for Section 4, which presents the observations of the LM and related phenomena that occur after the L-H transition, in particular the changes in the rotation of the MHD modes, the EF penetration threshold and its interpretation, and the TM seeding due to the sawtooth instability. In Section 5, the relationship of the presented results to other experiments and general implications are discussed, which is then summarized and concluded in Section 6.

## 2. Experimental setup and methods

### 2.1 The main experimental parameters, EF coil system

COMPASS was a compact-sized tokamak ($R = 0.56$ m, $a = 0.23$ m) with an open carbon divertor, an ITER-relevant plasma shape, and neutral beam injectors (NBI) that was operated in the Institute of Plasma Physics of the Czech Academy of Sciences until 2021 [19]. The reported experiments were performed in low $q_{95} \sim 3$ regime analogous to the ITER baseline [17], with toroidal field $B_t = 1.15$ T, plasma current $I_p = 230$ kA, and grad-B drift direction favorable to the L-H transition. In part of the experiments, one of the tangential on-axis NBI systems was used, delivering up to 400 kW power in 40 keV deuterium atoms injected in the co-$I_p$ direction. Calculation of the deposited power $P_{NBI}$ includes the beam duct losses [20].

The line-averaged density $n_e$, ranging from 1.5 to $9.5 \cdot 10^{19}$ m$^{-3}$, and the NBI power $P_{NBI}$ were the main control parameters that differentiated the scenarios used in this work. In the Ohmic experiments, reported in Section 3, $n_e \lesssim 4 \cdot 10^{19}$ m$^{-3}$. In the NBI-heated as well as Ohmic H-mode, the details of which are given in Section 4.1, $n_e \gtrsim 3.5 \cdot 10^{19}$ m$^{-3}$. Otherwise, all equilibria considered for LM occurrence within this study were quite similar except for the variation in $\beta_N$ (0.45 - 1) and plasma position $Z_{axis}$ (0 - 40 mm). The latter was used to control the input power threshold for the L-H transition, $P_{LH}$, through the height of the X-point above the divertor [19]. A typical plasma equilibrium shape and its main parameters just before an NBI-induced L-H transition, calculated by EFIT reconstruction constrained by the magnetic diagnostics, is shown in Figure 1. Position of the $q = 2$ surface was approximately at $\psi_N \sim 0.8$. The low $\beta_N$ in this work reflects the fact that only the first few ms of the H-mode are taken into account.

As a note on the used terminology, the Ohmic regime will generally refer to low confinement mode within this work unless explicitly termed Ohmic H-mode.

The $n = 1$ EF applied in this work was created by 3D ex-vessel coils located at the inboard side of the torus (HFS coils). The shape and position of the HFS coils is displayed in Figure 2 together with the resulting vacuum perturbation field at plasma separatrix. Although the HFS-originated EF (further referred to as HFS EF) also includes large non-resonant components, this work is focused on the resonant components only, which are conventionally assumed to have the dominant effect on plasma.

## 2.2 Diagnostics and data analysis methods

The objective of this paper is to examine the conditions for the onset of LM, which is characterized by a sudden change in the locked-mode detector signal caused by the growth of a non-rotating (or slowly rotating) 2/1 magnetic island. This section first presents the magnetic data, followed by other diagnostics used to study the conditions for LM.

*Magnetic data*

The $n = 1$ magnetic data are obtained from the set of ex-vessel saddle coils directly attached to the vacuum vessel and distributed over its surface in 24 poloidal and 4 toroidal positions (or 8 toroidal positions in the case of two LFS midplane rows). Their distribution is illustrated in Figure 3 [21], together with the coefficients used to combine the signals from one toroidal array into a single complex signal $B'_{r,n=1}$ representing the amplitude and toroidal phase of the $n = 1$ component of magnetic field normal to the vacuum vessel:

$$B'_{r,n=1} = -B'_{r,SE} - i.B'_{r,NE} + B'_{r,NW} + i.B'_{r,SW} \quad (1)$$

where $B'_{r,SE}$ represents the dB/dt data of a single saddle coil in the SE toroidal position. The array of sensors labeled XX2 (namely SE2, NE2, NW2, SW2), located just above the LFS midplane, was found to be the most suitable for observing the 2/1 mode due to its close proximity to the plasma. Its high signal-to-noise ratio allows tracking even a slowly rotating or locked mode. Therefore, all $n = 1$ data shown in this paper are from this location.

The locked mode detector utilizes the $n = 1$ normal magnetic field component $B_r$ obtained by integrating the $n = 1$ coil combination $B'_{r,n=1}$. Specifically, the absolute value of the resulting complex signal $B_r$ is used, which allows to observe the amplitude of any $n = 1$ mode regardless of its toroidal phase.

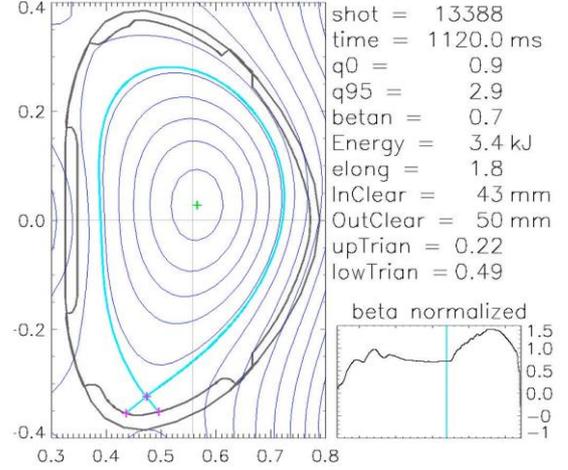

**Figure 1.** A typical plasma equilibrium shape and parameters prior to the L-H transition. The lower right inset shows the selected time point within a schematic time trace of $\beta_N$.

The dominant poloidal numbers of the observed modes were evaluated using the poloidal array of 24 Mirnov coils for representative time and frequency windows. It was found that, within the scenarios used in this work, the m/n=2/1 dominant mode can be routinely distinguished from the other $n = 1$ modes by the value of the phase shift between the saddle coils at two poloidal locations XX2 and XX21, above and below the LFS midplane. For rotating modes, the phase shift in the time domain is evaluated from the cross-coherogram of the corresponding $B'_{r,n=1}$ signals, which allows to routinely select the approximate region in the time and frequency domain that dominantly corresponds to the 2/1 mode. This information is then used to mask the spectrograms of individual $B'_{r,n=1}$ signals and calculate the time evolution of the 2/1 mode rotation frequency from the locations of the maximum amplitude in each time slice. Examples of this analysis are given in Section 4.2.

For locked and slowly rotating modes, whose amplitude evolves on a time scale shorter than the rotation period, the described Fourier analysis cannot be used and thus the concurrent modes at different frequencies are not separated. However, the instantaneous difference in the toroidal phase of the $B'_{r,n=1}$ signals from positions XX2 and XX21, which can be evaluated at the full time resolution, still gives an indication of whether the dominant mode has the $m = 2$ structure. This analysis will be used in Section 4.4.

*The other diagnostics*

The conditions for the onset of locked mode are quantified on a shot-to-shot basis in terms of the amplitude of the total resonant field overlap $\delta_{n=1}$, which represents the resonance of the EF with the core plasma, and the line-averaged density $n_e$, i.e. the two parameters expected to have the most significant effect on the EF penetration. The total $\delta_{n=1}$, resulting from both the applied error field (HFS EF) and the estimated intrinsic error field, is calculated by GPEC for an equilibrium at 2 ms before the LM onset. The estimation of the intrinsic EF is described in Section 3.1.

The line-averaged density, $n_e$, serves as the scaling parameter in this study, effectively discriminating the parameter space across all scenarios under examination. This parameter is obtained by integrating the plasma density profile measured by high-resolution Thomson scattering diagnostic (TS) [22], which is absolutely calibrated by Raman scattering and provides reliable data for a systematic scan over several years. To account for the limited time resolution of TS, an average over a 33 ms interval before the onset of LM is employed, which includes at least two plasma profile measurements by TS. Consequently, the error bars for the scaling points, as shown e.g. in Figure 5, include the standard deviation in this 33ms interval. The time evolution of $n_e$ is illustrated e.g. in Figure 4.

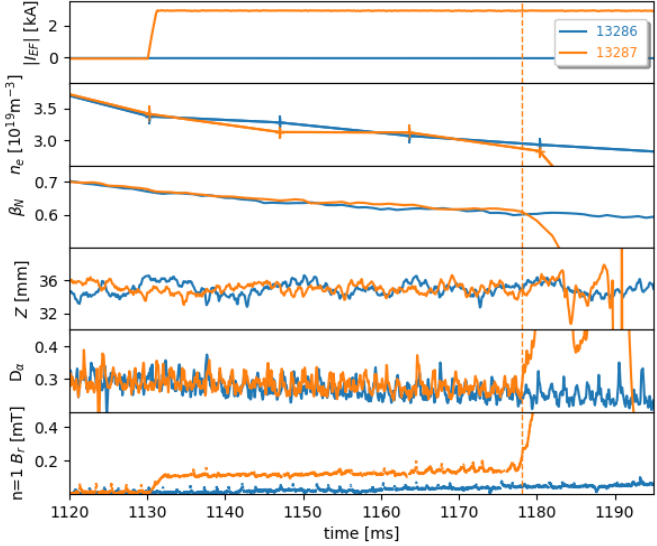

**Figure 2**. Approximate HFS EF coil segments as used in GPEC and their vacuum field normal component at the plasma separatrix.

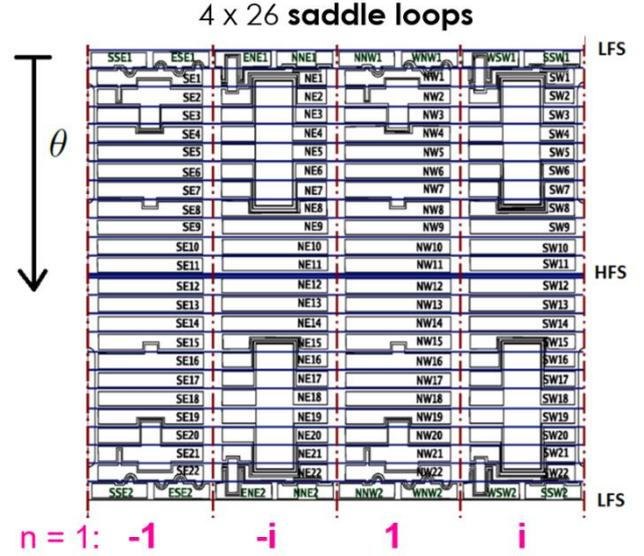

**Figure 3**. Schematic map of diagnostic saddle loops in poloidal and toroidal angles, and the coefficients used for the complex $n = 1$ combination in equation (1).

The hydrogen emission line $D_\alpha$ [23], sensitive to the plasma-wall interaction and thus the level of particle transport to the wall, serves as an indicator of the L-H transition and the ELM crash events (see Figure 6). Within this work, the time of the L-H transition is defined as the symmetry point of the $D_\alpha$ drop.

Plasma rotation profile measurement is not available, but the rotation frequency of the MHD modes, which is more relevant for the present study, is systematically obtained from the $n = 1$ magnetic signals $B'_{r,n=1}$ as presented in Section 4.2.

## 3. Critical density for EF penetration in Ohmic plasma

Since the line-averaged density $n_e$ is one of the most dominant factors in the multi-machine scaling [8], this part of the work aims to quantify the $n_e$-dependence of the HFS EF penetration threshold in Ohmic plasmas, which will provide a reference for the study of LM after the L-H transition.

The critical $n_e$ for EF penetration, $n_{e,\mathrm{crit}}$, was determined in a scenario where the density is ramped down during a constant pulse of HFS EF, a typical example of which is shown in Figure 4. The penetration event in discharge #13287 (orange) occurs at approximately 1178 ms, as indicated by the sudden increase in the $n = 1$ $B_r$ signal due to the nonlinear growth of the LM. The single data point obtained from the LM onset in a shot like the one shown thus defines, within the uncertainty of density measurement, $n_{e,\mathrm{crit}}$ for the penetration of the applied HFS EF of the given amplitude.

It can also be seen in Figure 4 that activation of HFS EF at 1130 ms induces only a linear response in the $n = 1$ $B_r$ signal (measured by a LFS saddle coil), while there is no apparent change in plasma confinement ($\beta_N$) or transport ($D_\alpha$). This is characteristic of the ideal-like plasma screening of the resonant EF up to its penetration.

This scenario was repeated for different levels of EF, which allowed us to explore the dependence of $n_{e,\mathrm{crit}}$ on the EF amplitude, quantified by the GPEC overlap metric $\delta_{n=1}$, as well as to explore the locked-mode-stable parameter space at $n_e > n_{e,\mathrm{crit}}$. Both types of information are summarized by individual colors of data points in Figure 5. The data points indicating the stable region (green) were evaluated from the minimum $n_e$ reached in shots where the EF penetration did not occur as well as from selected time points during the EF pulse prior to the penetration.

For reference, a few LM threshold data points were also obtained by ramping up the EF during nearly constant $n_e$. These points are compared to the main data set in Figure 5, but are not included in the fit.

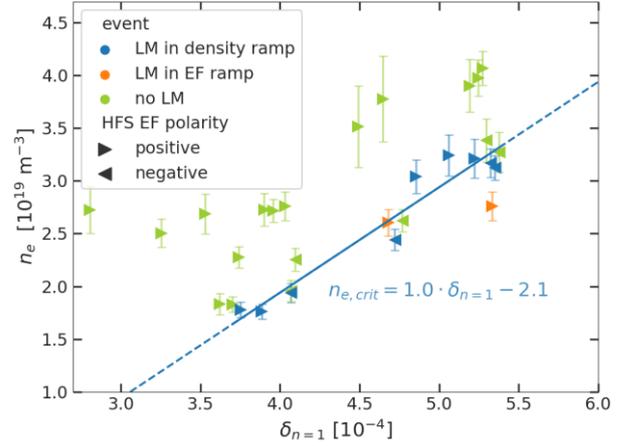

**Figure 5**. Line-averaged density $n_e$ versus EF overlap $\delta_{n=1}$ at EF penetration in Ohmic plasma during a ramp of $n_e$ (blue) or $\delta_{n=1}$ (orange), compared to the values of $n_e$ and $\delta_{n=1}$ at which the LM did not occur (green points). A linear fit through the blue data points (blue line, dashed for extrapolation) represents the boundary of the locked-mode stable region for Ohmic plasma in COMPASS. Penetration of two polarities of the applied EF (different symbols) occurs at similar $n_e$ and $\delta_{n=1}$ due to the inclusion of the intrinsic error field.

**Figure 4**. The Ohmic plasma density ramp-down scenario in two identical shots, except for the presence (orange) or absence (blue) of the applied HFS EF. From top to bottom, the time traces of the EF coil current, line-averaged density, normalized $\beta$, vertical position, $D_\alpha$ radiation, and locked-mode detector $B_r$ (integrated $n = 1$ field) are shown. The dashed vertical line marks the EF penetration (orange).

Figure 5 shows a clear positive correlation between $n_{e,\mathrm{crit}}$ and $\delta_{n=1}$. Although the functional form of the dependence cannot be obtained reliably from the given data, a linear fit is able to separate the stable and unstable regions of the locked mode and is also consistent with the linear density scaling observed in separate experiments in JET, COMPASS-D, and DIII-D [24]:

$$n_{e,crit} = 1.0 \cdot \delta_{n=1} - 2.1 \qquad (2)$$

However, the finite offset of the fit indicates that this linear scaling is not valid below the density range covered in this experiment.

The presented behavior of HFS EF penetration in Ohmic plasma is apparently consistent with the experience from the existing LM scaling experiments with the more typical LFS EF. Furthermore, preliminary results of another COMPASS experiment show that there is no significant difference in the EF penetration threshold between a stabilized NBI-assisted H-mode and Ohmic plasma, and between the HFS- and LFS-originated EF, which will be further investigated in a forthcoming paper focusing on the agreement or disagreement of our data with the multi-machine scaling. It can thus be concluded that the presented linear scaling (2) provides a sufficient reference for the L-H transition study presented in Section 4.3.

## 3.1. Intrinsic EF

The presence of the intrinsic EF leads to an asymmetry in the penetration of the opposite toroidal orientations of the applied HFS EF. To account for this in the analysis, an additional experiment was performed to estimate the amplitude and phase of the resonant overlap of the intrinsic $n = 1$ EF using the so-called compass scan (explained in [25] for DIII-D, previously used in many tokamaks, e.g. JET [24], C-MOD [25a], MAST [25b], NSTX [25c]). The LFS 3D coils were used, which, in contrast to the HFS coils, were available in 4 different toroidal phases.

The magnitude of the identified intrinsic EF was estimated to $\delta_{n=1, IEF} = 2.3 \cdot 10^{-4}$. With additional assumptions on how the coupling of the intrinsic EF is affected by the small variations in equilibrium, which is important when comparing data from different sets of experiments and is reflected by the horizontal error bars shown later in Figure 11, the intrinsic field is included in the total EF overlap value $\delta_{n=1}$ throughout this paper. A reasonable agreement is shown in Figure 5 between the data points with positive and negative applied HFS EF polarity (indicated by different symbols) after including the intrinsic EF.

A more detailed analysis of the intrinsic error field in COMPASS, including the results of the compass scan, will be covered in a future paper.

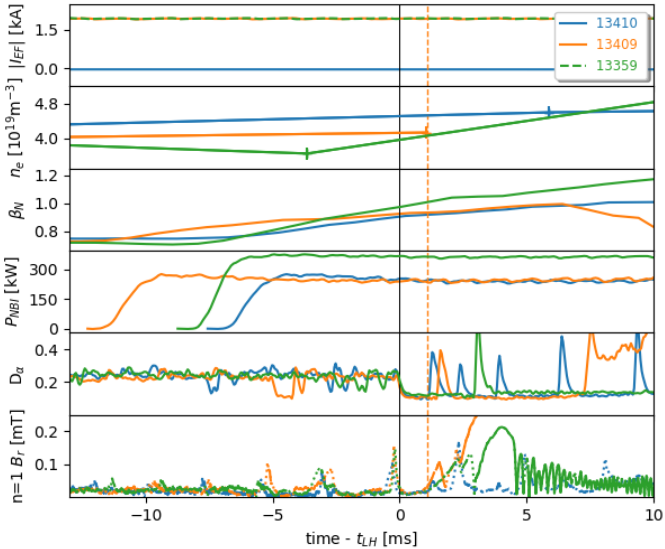

**Figure 6.** NBI-assisted L-H transition in a reference shot with no EF applied (blue) and in cases with disruptive (orange) and non-disruptive (green) penetration of applied HFS EF. The time evolution relative to the time of the L-H transition $t_{LH}$ (vertical solid line) is shown for the EF coil current, line-averaged density $n_e$, normalized $\beta$, NBI power, $D_\alpha$ radiation, and absolute value of complex $n = 1$ magnetic field $B_r$, respectively. The locked mode onset (vertical dashed line) in the disruptive shot is identified from the nonlinear growth of $B_r$.

## 4. Observation of LM after the L-H transition

In the previous section, it was shown that in the COMPASS Ohmic plasma, disruptive LMs due to the HFS EF occur systematically as a function of $n_e$. However, when the HFS EF is present during the L-H transition, this predictability is lost and LMs occur in a range of parameters previously identified as stable. This is demonstrated in the experimental sequence discussed in Section 4.1, which shows the LM as an apparent consequence of the L-H transition. Natural mode rotation is discussed in Section 4.2 and provides an explanation for this unexpected LM appearance after the L-H. The parametric space for LM occurrence is then related to the reference LM stability boundary for Ohmic plasma in Section 4.3, where a new threshold for EF penetration during the L-H transition is defined. For disruptive LMs occurring below this penetration threshold, an alternative triggering mechanism based on TM seeding by sawtooth is proposed in Section 4.4.

### 4.1. Scenarios with HFS EF during the L-H transition

The data set under study includes three L-H transition scenarios, each used in a different $n_e$ window. The H-mode had a non-stationary ELM-free character when it was entered at $n_e > 4.5 \cdot 10^{19}$ m$^{-3}$, and ELMy otherwise. In all cases, only the L-H transitions occurring during the stationary phases of Ip and plasma shape were considered, with the EF coil current held constant for at least 20 ms prior to the L-H transition.

The first L-H transition scenario is shown in Figure 6, where the NBI was activated in the Ohmic phase, causing the input power to exceed the $P_{LH}$ threshold. The 3 shots in different colors are aligned by the time of the L-H transition $t_{LH}$, defined by the drop of the $D_\alpha$ emission. ELM peaks in $D_\alpha$ are present after the transition and a dithering phase can be seen before it. Nevertheless, there is a clear difference between the L- and H-mode visible on a short time scale. The limited density range at the L-H transition and a moderate $P_{NBI} < 300$ kW typically resulted in a stable ELMy H-mode in COMPASS, avoiding the high-energy disruptions typical of the NBI-heated non-stationary ELM-free H-mode. The vertical plasma position Z was kept constant in this scenario, mostly at Z = 20 mm.

Figure 6 illustrates two different outcomes resulting from the application of the same HFS EF in this scenario, represented by reference shot #13410 (blue). The first case with HFS EF (#13409, orange) resulted in a disruptive LM after the L-H transition, with EF penetration at 1 ms indicated by the orange vertical line. In the second case (#13359, green), a similar EF penetration can also be observed, with the same delay of 1 ms after the L-H transition. However, the resulting magnetic island is eventually accelerated and stabilized in #13359, most likely due to the higher NBI torque and power in this shot ($P_{NBI} > 370$ kW) compared to the group of shots with disruptive LM ($P_{NBI} < 320$ kW). In addition, the bottom two plots show that the growth of the $n = 1$ signal was disrupted by the ELMs at 1.6 ms in #13409 and at 3 ms in #13359.

The second scenario used is an Ohmic H-mode triggered by the vertical plasma position $Z$, as shown in Figure 7. Here, the $P_{LH}$ was gradually decreased by a downward ramp of the X-point height [19] in high-density Ohmic plasma, which allowed a sharp L-H transition on the high-density branch of the $P_{LH}$, typically without any dithering. The resulting non-stationary ELM-free H-mode was usually terminated by a non-disruptive H-L transition. The regular peaks in the $n = 1$ magnetic field, most clearly visible before the L-H transition in this figure, represent the signature of sawtooth crash events. This signature will be defined as ST mode in this paper and discussed further in the following sections. Its peaks are aligned between shots because the L-H transition itself is typically triggered by sawtooth in COMPASS [26, 23].

Again, several possible consequences of the L-H transition are shown as an example in Figure 7. In #18161 (orange), the penetration of the applied EF occurs 1.5 ms after the L-H transition. In #18162 (pink), despite no applied EF, a nonlinear growth of the $n = 1$ field starts at 3.5 ms, which is interpreted as locking of a TM seeded by a sawtooth crash. Non-disruptive TM seeding is also observed in reference case #18151 (blue). These types of events are further elaborated in Section 4.4.

The third type of the L-H transition is also Ohmic, but in a constant $Z$ scenario presented in Figure 6, except that no NBI was applied. In this case, the Ohmic H-mode was entered spontaneously, most likely as a result of variation in gas puff and/or density near the optimum for $P_{LH}$. These cases provide a useful bridge between the Ohmic and NBI-assisted scenarios at moderate densities. As in the second scenario, the resulting H-modes are mostly ELM-free, although one case of ELMy H-mode is included for comparison.

*4.2. Rotation changes after the L-H transition*

The analysis of the magnetic fluctuations provides useful information on the rotation frequency of the MHD modes. The mode frequency, interpreted in this paper as the toroidal mode rotation, serves directly as the relevant parameter for the EF penetration in the theoretical models mentioned in Section 1. The advantage of the direct measurement of the mode rotation, in contrast to the common approach based on ion rotation data, is that it does not depend on the location of the rational surface, nor on assumptions about the mode propagation with respect to ion or electron fluids, nor on the diamagnetic frequencies.

Two changes in mode rotation associated with the L-H transition in COMPASS can be systematically inferred from the $n = 1$ magnetic data, both of which have implications for the occurrence of LM and disruptions. First, the rotation of the 2/1 tearing mode (TM) decays rapidly from its L-mode level of about 10 kHz in the counter-$I_p$ direction, reverses direction to co-$I_p$, and then typically remains at about 1 kHz throughout the H-mode. Second, the rotation of the mode associated with sawtooth instability, which appears as a precursor and/or postcursor to the sawtooth crashes, also changes direction from counter-$I_p$ and then is significantly accelerated in the co-$I_p$ direction when the external torque from the NBI is present.

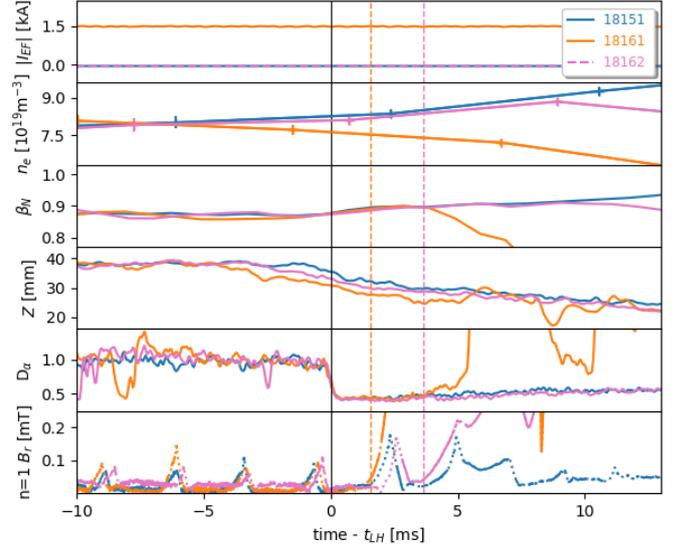

**Figure 7.** Ohmic L-H transition triggered by vertical X-point movement in a reference shot (blue) and in two shots with locked mode onset (vertical dashed lines), either due to penetration of applied HFS EF (orange) or due to locking of a tearing mode seeded by sawtooth without applied EF (pink). The figure is organized in the same way as Figure 6, except that the vertical plasma position $Z$ is shown instead of the NBI power.

*2/1 tearing mode rotation*

The rotation of the 2/1 TM, localized at about $\psi_N \sim 0.8$ according to the equilibrium reconstruction, is well visible in the spectrogram of the $n = 1$ component if the mode is strong enough to be dominant in its frequency range. In our experiment, this was typically the case for the Ohmic plasma and the L-mode, but not always for the H-mode. The mode is also stronger at higher $n_e$ as expected for a TM, while at $n_e \lesssim 4 \cdot 10^{19}$ m$^{-3}$ it tends to disappear even before the L-H transition.

Two examples of a well-resolved 2/1 mode, for the Ohmic and the NBI-assisted L-H transition scenarios, are shown in the spectrograms in Figure 8, where the time and frequency windows with m=2 dominant signal are highlighted by two colors for the two rotation directions. The masks for these windows were calculated from the phase shift of the $B'_{r,n=1}$ signals at two nearby poloidal locations as described in Section 2.2. The frequency of the 2/1 mode in each time slice is calculated from the location of the maximum amplitude within the frequency window and is plotted by solid lines in Figure 8. False detections are minimized by several measures, such as removing outlier frequency jumps (above 10 kHz/ms) and requiring that the mode frequency can be detected in more than 50% of spectrograms of $n = 1$ combinations from different toroidal and poloidal locations around the outer midplane.

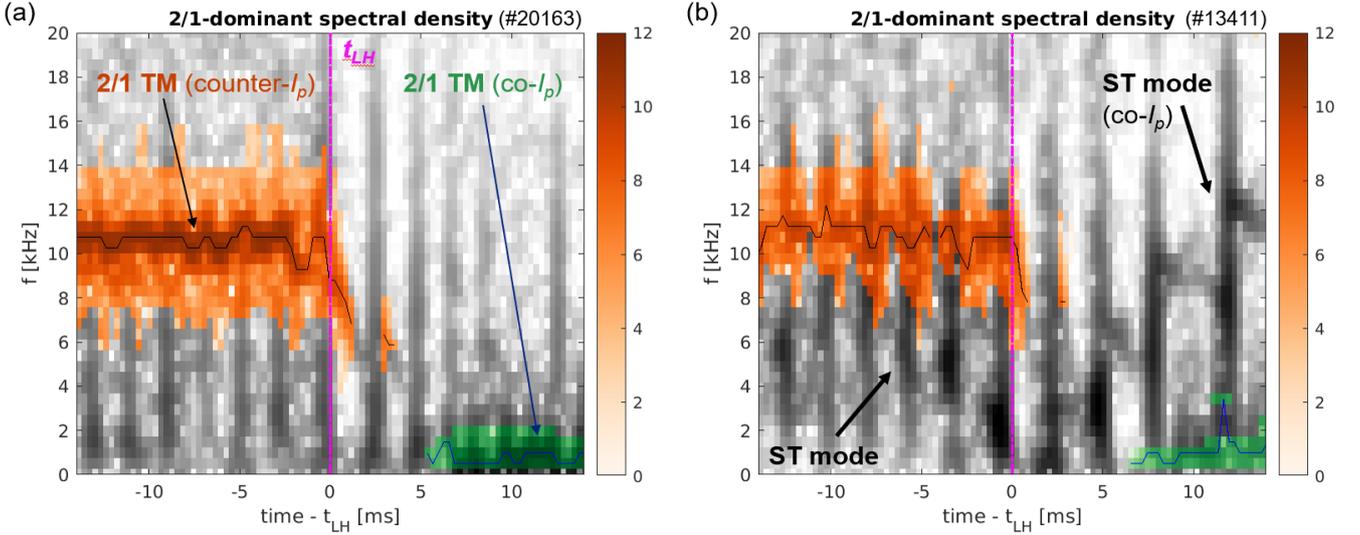

**Figure 8**. Spectrograms of the real part of the $n = 1$ signal $B'_{r,\,n=1}$ in two L-H transitions with no EF applied, (a) with an Ohmic L-H transition and (b) with a 400 kW NBI activated 7 ms before the L-H transition, show the typical time evolution of a 2/1-dominant mode. The frequency windows highlighted by color palettes correspond to this mode rotating in the counter-Ip direction (orange palette) and co-Ip direction (green palette). The frequency of the maximum amplitude within these windows is represented by solid lines (black and blue, respectively).

The 2/1 rotation decay and reversal occur during the first 5 and 7 ms of the H-mode, respectively, in the two cases shown in Figure 8. Its full time evolution from L-mode level to H-mode level rotation cannot typically be resolved in a single discharge, partly because the amplitude of the 2/1 mode is significantly reduced around the L-H transition, partly because of overlap with other MHD activity (mainly sawtooth).

The sparsity of the 2/1 rotation data in individual shots can be improved by conditional averaging over groups of similar shots. This is illustrated in Figure 9, where the variations in frequency between the shots in each group are represented by shaded areas. The very small variation up to 1.25 ms after the L-H transition (dotted line) shows that the general behavior of the 2/1 rotation is reproducible and the $t_{LH}$ from the $D_\alpha$ radiation is well defined in these groups. The high uncertainty in the later time is due to the fact that the 2/1 mode is detected in fewer shots and, in cases without LM, it reaches the rotation reversal at different times.

In the ELM-free cases, the presence of applied HFS EF has no significant effect on the decay of the 2/1 mode rotation after the L-H transition, as seen in Figure 9. Also, the discharges where LM was developed show the same time evolution of the 2/1 rotation up to the point of locking. It is therefore suggested that the observed behavior corresponds to the natural rotation of the 2/1 mode after the L-H transition, caused by the change in confinement itself. There is also no evidence that the 2/1 rotation is affected by the high power NBI, except for its slight acceleration in the co-$I_p$ direction after 20 ms of H-mode, as shown by solid lines in Figure 10.

For the transitions to the ELMy H-mode, which occur at $n_e < 4.5 \cdot 10^{19}$ m$^{-3}$ (see Figure 11 for the parameter space), there are not enough 2/1 mode rotation data to conclude whether the applied EF or the NBI has a significant effect on the mode rotation or not. A clear rapid decay of the mode rotation, similar to the ELM-free cases shown in Figure 9, is only seen in the ELMy cases with LM. In the ELMy H-mode without LM, the decay of the 2/1 mode rotation cannot be followed, but in many cases the 2/1 mode with reversed rotation direction appears after 10 or 20 ms of H-mode, indicating that the rotation decay must have taken place.

*ST mode (sawtooth mode) rotation*

The other MHD activity in the spectrograms in Figure 8 is dominated by the $n = 1$ component of the sawtooth instability, referred to in this paper as the "ST mode". It appears as intermittent bursts of the precursor/postcursor mode typically associated with the 1/1 resonant surface. It is difficult to precisely assess the frequency and poloidal structure of the ST mode due to the short time scale (which also implies its broadband appearance in the spectrogram), but there are indications that the structure is quite complex, probably due to the coupling of the 1/1 mode to higher m modes. The internal kink associated with sawtooth has also been observed by the soft X-ray diagnostic in COMPASS, both before and after the sawtooth crash [26].

From the spectrogram of a high power NBI shot shown in Figure 8 (b), the frequency of the ST mode was approximated from the maximum intensity within its respective frequency windows and is shown by red symbols in Figure 10. Unlike the 2/1 mode, the ST mode rotation is significantly shifted in the co-$I_p$ direction by the acceleration of the core plasma by the co-injected NBI. This results in a clear frequency separation between the ST and 2/1 modes during the NBI-assisted H-mode. Furthermore, the sawtooth periods and the intensity of the crashes gradually increase as a consequence of the NBI [27, 28].

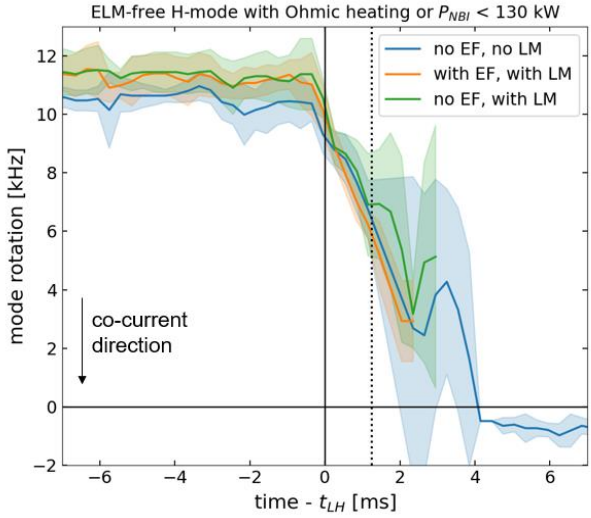
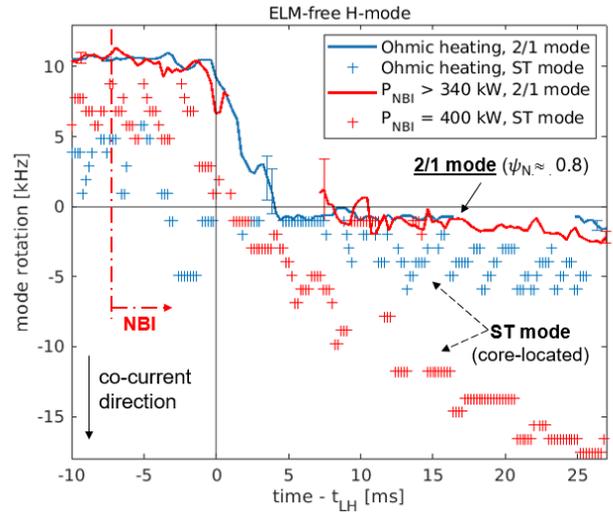

**Figure 9**. Time traces of the 2/1 mode rotation frequency after the L-H transition in ELM-free H-mode scenarios with either Ohmic or low-power NBI heating show a consistent rapid decrease at the transition. The shaded areas represent the standard deviation of mode rotation among discharges in each group, which is particularly small until 1.25 ms after the L-H transition (dotted line).

**Figure 10**. Averaged 2/1 mode rotation over two groups of shots (solid lines) and approximate sawtooth mode rotation in two shots #18151 and #13411 (plus symbols), for Ohmic (blue) and high-power NBI (red) ELM-free H-mode, respectively. NBI significantly affects the sawtooth mode, but not the 2/1 mode.

Neither of the above is the case for the Ohmic H-mode in Figure 8 (a), where the magnitude of the ST mode is generally smaller, and for a certain time interval around the L-H transition, the mode rotates at a very low frequency in either the co-$I_p$ or counter-$I_p$ direction. In some cases, the direction of rotation of the ST mode can only be distinguished by the time evolution of the amplitude and phase of the integrated magnetic data, as shown in Figure 13.

As shown in Figure 10, the ST mode rotation frequency in the Ohmic H-mode (blue symbols) tends to overlap with the 2/1 mode rotation (blue line) for a short time interval of about 5 to 10 ms after the L-H transition when the rotation of both modes is close to zero. It is at this moment that the 2/1 mode signal typically becomes stronger, in what we consider to be a TM seeding event. This will be discussed in Section 5.2 with a proposed explanation by 3-wave coupling at low differential rotation.

### 4.3. EF penetration after the L-H transition

The conditions for the onset of the disruptive LM after the L-H transition in COMPASS, in terms of density $n_e$ and EF overlap magnitude $\delta_{n=1}$, are indicated by red data points in Figure 11, one for each shot from the examined data set. This LM clearly occurs well within the stable region for Ohmic plasma characterized by the critical density $n_{e,\mathrm{crit}}$ (blue line and dots) defined in Figure 5. As discussed in Section 3, this $n_{e,\mathrm{crit}}$ is consistent with LM thresholds in other devices as well as in the stabilized NBI-heated H-mode in COMPASS. After the L-H transition, in contrast, LM is observed even at much higher $n_e > 4 * n_{e,\mathrm{crit}}$ for the given $\delta_{n=1}$, with no apparent upper density limit, and even at such low $\delta_{n=1}$ where no LM is observed in Ohmic plasma. Shots with no disruption after the L-H transition are included in the figure for reference (green dots) and to indicate the disruption probability. At first glance, the probability increases with $\delta_{n=1}$ and does not depend on $n_e$.

Figure 11 includes all 3 types of the L-H transition, showing no significant difference in disruption probability within the range and scatter of the available data. In particular, Ohmic L-H transitions triggered by the change in plasma position correspond to the data points at $n_e > 7 \cdot 10^{19}$ m$^{-3}$, and the remaining two types are distinguished by the markers indicating the levels of NBI power.

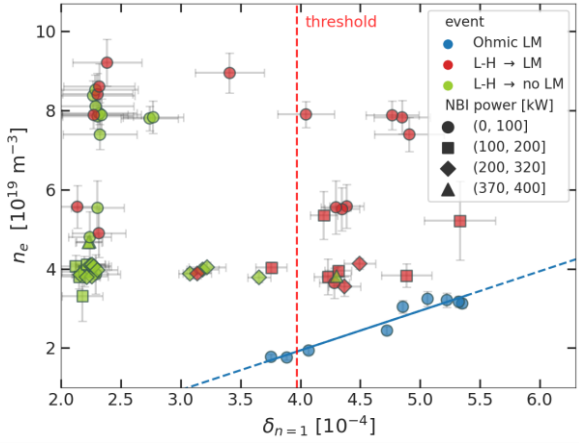

**Figure 11**. Line-averaged density $n_e$ versus EF overlap $\delta_{n=1}$ at the onset of a disruptive locked mode after the L-H transition (red), compared to the cases with no disruption after the L-H transition (green), shows a high disruption probability in the parameter region that is stable for the locked mode in Ohmic plasma, represented by the fitted points (blue) taken from Fig. 5. The dashed red line represents the assumed threshold for EF penetration after the L-H transition. The horizontal bars reflect the uncertainty in the different intrinsic error field coupling between individual experiments. The levels of NBI power are indicated by different symbols.

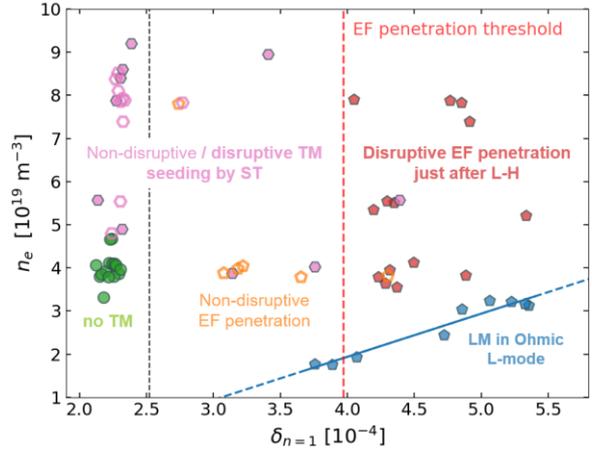

**Figure 12**. The same data in the same parameter space as in Figure 11, but the artificial seeding of locked modes by EF penetration (red pentagons) is distinguished from the spontaneous locking of a sawtooth-seeded tearing modes (pink hexagons). The empty symbols represent non-disruptive tearing modes seeded by either sawtooth (pink hexagons) or EF penetration (orange pentagons) after the L-H transition. The points to the left of the black dashed vertical line are without applied EF. The red dashed vertical line is like in Figure 11.

*Threshold for EF penetration*

A threshold for EF penetration after the L-H transition is observed at the EF overlap magnitude $\delta_{n=1} \cong 4 \cdot 10^{-4}$, as indicated by the dashed red line in Figure 11. This threshold represents two distinct changes in plasma behavior. First, the disruptive LM becomes very likely to appear after the L-H transition at $\delta_{n=1}$ above the threshold, occurring in all shots except the one with higher NBI power (#13359). Second, the observed LM onsets will have a different character and dynamics, allowing to distinguish the origin of the TM between the one artificially seeded by the penetration of the applied EF and the one spontaneously seeded by the sawtooth activity, as presented in Section 4.4. From Figure 12, which shows the distribution of individual TM events over the data set, it is clear that the disruptive EF penetration after the L-H transition is indeed observed only above the threshold.

It is of particular interest that the threshold for EF penetration after the L-H transition does not appear to scale significantly with density (Figure 12), contrary to existing scaling laws for steady-state plasma regimes. However, within the available data set, it cannot be excluded that the density dependence is canceled out by the dependence on NBI power.

*Consideration of 2/1 mode rotation*

The decrease in the 2/1 natural frequency after the L-H transition consistently explains the particularly low EF penetration threshold compared to steady-state conditions. The timescale of the two events is strongly correlated and much shorter than the L-mode energy confinement time in COMPASS ($\tau_E > 7$ ms) [29]. Above-threshold EF penetration typically occurs 1 - 1.5 ms after the L-H transition. At this time, the 2/1 frequency drops to approximately 50% of its L-mode level. Several shots exhibited fast ELMs or oscillations in $D_\alpha$ that may correspond to limit cycle oscillations [30], delaying both the frequency drop and EF penetration. Therefore, the 50% reduction in the 2/1 natural frequency is likely sufficient to allow penetration of an EF above the threshold in this case.

One would expect the plasma to become even more sensitive to the EF as the 2/1 rotation frequency continues to decrease. This is indeed observed in several shots where a weak applied EF penetrates with a longer delay of 3.5 - 10 ms after the L-H transition. However, the penetration of this weak EF does not lead to disruptive LMs, but rather to rotating modes that are eventually stabilized (orange empty symbols in Figure 12).

*4.4. Locking of seeded TM after the L-H transition*

According to the analysis presented in this section, the locking of sawtooth-seeded TMs is the cause of all disruptions after the L-H transition at $\delta_{n=1} < 4 \cdot 10^{-4}$. This is indicated in Figure 12. To distinguish between the locking of a pre-existing sawtooth-seeded TM (filled magenta symbols) and the penetration of static external EF (red symbols), we present a detailed analysis of the amplitude and phase of the $n = 1$ signal. In addition, sawtooth-seeded TMs themselves are also observed in many non-disruptive shots (open magenta symbols) at $n_e > 4.5 \cdot 10^{19}$ m$^{-3}$. The proposed mechanism for why the seeded TM only locks in a fraction of cases is given in the second part of this section. Finally, there is a considerable number of shots where no TM is detected (green symbols) at $n_e \lesssim 4.5 \cdot 10^{19}$ m$^{-3}$. This is likely related to the presence of an external torque that induces a differential rotation between the ST and 2/1 modes (shown in Fig. 10), removing the conditions for resonant TM seeding.

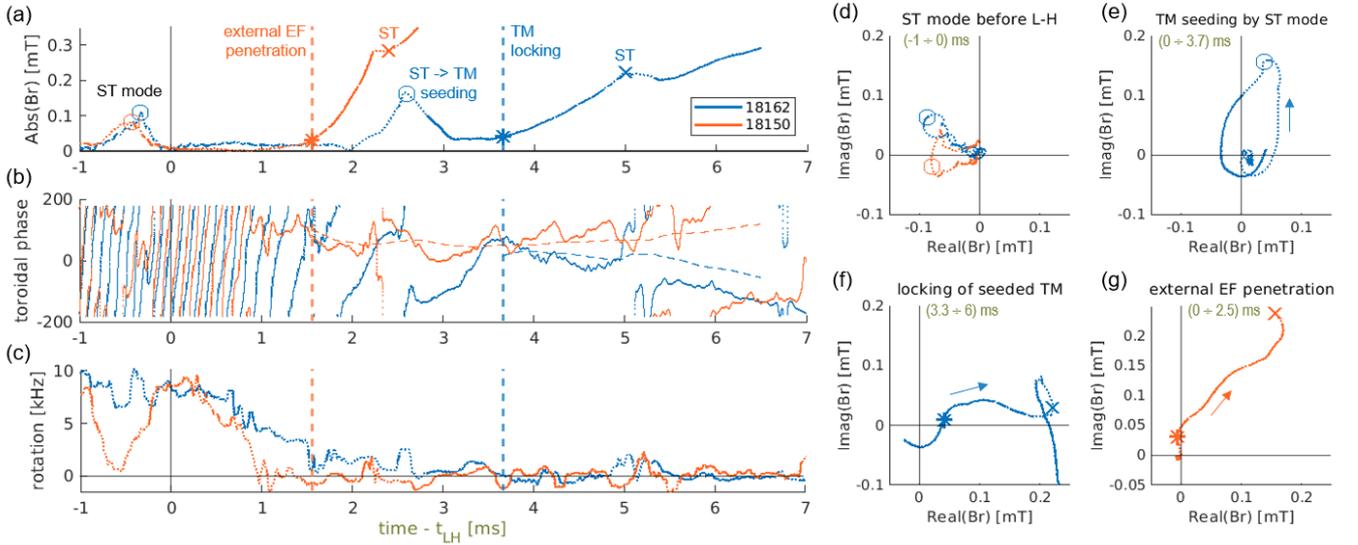

**Figure 13**. Comparison of the integrated $n = 1$ magnetic field component $B_r$ in two shots showing a locked mode after the Ohmic L-H transition caused either by penetration of the applied EF (orange) or by locking of a sawtooth-seeded tearing mode (blue). The time evolution relative to the time of the L-H transition $t_{LH}$ is shown for the amplitude (a), toroidal phase (b), and toroidal rotation (c) of the $n = 1$ field $B_r$. The shape of the $B_r$ trajectory in the complex plane, representing a view from above, is given in parts (d-g) for individual time intervals indicated. The signal with 2/1-dominant structure corresponds to the solid lines in parts (a), (c-g), while the dotted lines are for other $n = 1$ modes. Star, circle, and cross markers in parts (d-g) correspond to those in part (a) and indicate the onset of locked mode growth, ST crashes before the onset, and the first ST crash after it, respectively. The toroidal phase in (b) is calculated from both the integrated (dashed) and non-integrated (solid) $B'_{r,n=1}$ signal.

As a note on the terminology used, we do not consider the seeded TMs to be neoclassical, since the modes are stabilized unless their growth is supported by locking to the wall or external EF. Apparently, the seeded island size is not sufficient for neoclassical growth in this range of $\beta_N \sim 1$. Furthermore, it should be noted that the available diagnostics do not allow for distinguishing between the kink and tearing character of the modes in this experiment. The transition from a kink to a tearing mode at the 2/1 resonant surface can occur without being detected in the amplitude of the magnetic signal [31]. We assume that the tearing character of the mode becomes dominant once the observed mode has an $m = 2$ dominant structure.

*Identification of the locking of TM seeded by sawtooth*

The detailed time evolution of the $n = 1$ normal magnetic field $B_r$ in Figure 13 allows to observe the difference between the case of locking of a pre-existing sawtooth-seeded TM (blue) and the case of external EF penetration (orange), which will be addressed in the following paragraphs. The complex $B_r$ signal is obtained by integrating the complex $n = 1$ coil combination $B'_{r,n=1}$ introduced in Section 2.2. The time intervals during which the signal has a 2/1-dominant structure are obtained by requiring that the instantaneous difference in the toroidal phase of the $B'_{r,n=1}$ signals from two poloidal positions falls within a given range, similar to obtaining the masks for the spectrograms in Figure 8. Here in Figure 13, the mask corresponding to the 2/1 mode is represented by the solid lines in parts (a) and (c-g), while the dotted lines represent all other $n = 1$ modes. The method used to analyze $B_r$ will be elaborated in a separate publication. Here we focus only on what is needed to compare the two discharges shown.

The time trace of the $B_r$ amplitude (a) in Figure 13 represents a locked-mode detector. The peak around -0.5 ms corresponds to the ST mode introduced in Section 4.2 and also visible in Figure 7. The toroidal phase (b) and rotation frequency (c) provide a qualitative indication of the toroidal rotation of whichever $n = 1$ mode is currently dominant. The rotation drop of about half during the first ms of the H mode from near the L-mode level is consistent with the evolution of the 2/1 mode rotation in the spectrograms in Figure 8. Comprehensive information about the amplitude and phase of the $n = 1$ signal is contained in the shape of its trajectory in the complex plane (d-g), but at the cost of missing the time axis. To improve readability, the trajectory is shown at shorter intervals, with the markers from part (a) indicating the crucial time points. For example, part (d) illustrates that the ST mode peaks between -1 and 0 ms have a nearly constant phase in each of the shots and that the signals return to the origin after the peaks.

The case of external EF penetration (orange) in Figure 13 shows a typical picture of a LM in a shot with EF above the $\delta_{n=1} \cong 4 \cdot 10^{-4}$ threshold. First, the toroidal phase becomes stationary when the rotation is abruptly lost at 1 ms. Second, the signal becomes 2/1-dominant at a very low amplitude.

Finally, the nonlinear growth of the mode amplitude starts at about 1.5 ms (star symbol), which is well before the first sawtooth crash in the H-mode at 2.4 ms (cross symbol). This crash then only perturbs the mode growth, as does the crash at 5 ms for the blue curve. The trajectory in (g) shows the non-rotating 2/1 mode growth from near zero.

In the case of spontaneous TM seeding and subsequent locking (blue), the events occur in a different order. First, the 2/1-dominant signal appears with a finite amplitude at 2.8 ms, just after the first peak of the ST mode after the L-H transition. This is also illustrated by the signal trajectory (e), where the decay of the ST mode peak ends in a rotating TM of constant amplitude instead of returning to the origin. Second, the toroidal phase becomes stationary at 3.7 ms (star symbol) through a gradual evolution all the way from the L-H transition. At the same time, the nearly locked TM starts to grow, as shown also in part (f).

*Mechanism for locking the pre-existing TM*

Figure 12 shows that TM seeding by sawtooth occurs in a significant portion of the dataset. However, only about 30% of them develop into a disruptive LM, despite the slow rotation shown in Figure 13. The proposed mechanism for locking a seeded TM depends on the relative timing of TM seeding and the natural mode rotation reversal from counter-$I_p$ to co-$I_p$. In particular, mode locking occurs only if the rotation reversal occurs after the TM is seeded. This mechanism is independent of the applied EF, as is the evolution of the natural mode rotation from counter-$I_p$ to co-$I_p$ in ELM-free H-modes (see Figure 9).

This implication is experimentally confirmed by the fact that the disruptive and non-disruptive shots differ in the direction of mode rotation at the onset of a 2/1-dominant signal, i.e., at TM seeding. The case shown in Figure 13, where the direction at TM seeding is counter-$I_p$, is generally representative of all disruptive cases. Conversely, if the direction at TM seeding is already co-$I_p$ (i.e., TM seeding occurred after the reversal), the mode continues to slowly accelerate and eventually disappears. Finally, the random distribution of disruptive and non-disruptive cases of TM seeding in Figure 12 can be explained by the natural scatter in the timing of sawtooth crashes and 2/1 mode rotation reversals in individual shots.

In addition, Figure 12 shows two outlier shots with disruptive sawtooth-seeded TMs in the presence of external torque at $n_e < 4.5 \cdot 10^{19}$ m$^{-3}$ and $\delta_{n=1} < 4 \cdot 10^{-4}$. These TMs were seeded already as non-rotating, by a ST mode that had most likely been locked to the external EF prior to seeding.

**5. Discussion**

In the previous sections, two distinct physical mechanisms have been identified as the cause of the appearance of LM after the L-H transition: the penetration of the applied EF and the locking of the sawtooth-seeded TM. These mechanisms are further discussed in Sections 5.1 and 5.2. Finally, the potential implications and extensions of this work to other devices are discussed in Section 5.3.

*5.1 EF penetration threshold vs. the 2/1 mode rotation*

In Section 4.3, it was reported that the HFS EF penetration consistently occurs 1 - 1.5 ms after the L-H transition in COMPASS in all three scenarios considered. We attribute this to a 50% reduction in the natural rotation frequency of the 2/1 mode compared to its L-mode level. While this would be expected within the current understanding of the penetration process, it demonstrates the limitations of universally applying the global parametric EF penetration scaling (e.g., [8]) to the entire discharge. As we have shown, neglecting the transient phases during which the island rotation changes significantly in favor of correlating the natural mode rotation with the global plasma parameters results in disruptions at unexpectedly low EF levels.

Experiments in other tokamaks have also shown that the mode rotation frequency can spontaneously evolve and affect the $n = 1$ EF penetration threshold in unexpected ways. In addition to the already mentioned edge rotation braking during the first 30 ms of Ohmic H-mode in COMPASS-D [14], a similar phenomenon was recently observed when increasing the density in circular Ohmic plasma in J-TEXT [11]. There, the natural mode rotation was estimated from the impurity rotation by an empirical relation and shown to explain the puzzling non-monotonic dependence of the $n = 1$ penetration threshold on both the line-averaged density and the local density at $q = 2$.

*5.2 Resonant TM seeding by 3-wave coupling*

Figure 11 shows that disruptive LM can occur in shots with below-threshold EF, including those with only intrinsic EF, with an apparent disruption probability of about 30% in the absence of external torque. As explained in Section 4.4, all of these disruptions can be attributed to TM seeding by sawtooth. When non-disruptive cases are included, TM seeding by sawtooth is consistently observed after all studied transitions to the Ohmic ELM-free H-mode, except for cases where EF penetration occurs earlier (typically before the first sawtooth crash).

It may seem puzzling that TM seeding by sawtooth occurs more readily near the L-H transition, at $\beta_N \sim 1$, and with significantly smaller sawtooth crashes than during a fully developed H-mode. Under these conditions, the profile change after the sawtooth crash is unlikely to cause TM growth. For instance, TM seeding was reported at $\beta_N \sim 0.8$ in JET [32] only due to a large crash after a particularly long sawtooth period, by no means during regular discharge behavior. However, the 2/1 TM seeding observed after the L-H transition in COMPASS can be explained by its coupling to the sawtooth precursor mode via its $n = 1$ component. This process, known as resonant TM seeding, occurs due to the lack of rotation shear between the coupled MHD modes [33]. Figure 10 shows the absence of this shear shortly after the L-H transition in the Ohmic ELM-free H-mode scenarios.

The observed phenomenon is likely related to the disruptive 2/1 mode seeding through 3-wave coupling of the 1/1 sawtooth precursor to higher m/n tearing modes, recently demonstrated in an ITER baseline experiment in DIII-D [34] at $\beta_N \sim 1.8$. In this case, the good diagnostic coverage and longer time scales in DIII-D allowed for the observation of TM seeding during the growth of the sawtooth precursor mode, 14 ms before the sawtooth crash itself. The coupling between the ST and 2/1 modes in COMPASS may also be explained by the conventional toroidal coupling between the 1/1 and 2/1 modes, which is a special case of 3-wave coupling through the 0/1 perturbation due to Shafranov shift [35]. This type of TM seeding is closely related to the differential rotation between the involved modes, which produces a stabilizing term for the seeded mode. In DIII-D, TM seeding can be avoided if a differential rotation of at least 1 kHz is maintained [36]. This finding is in qualitative agreement with the observations made in COMPASS.

### 5.3 Implications for other devices

As reported in this work, the disruptive LM can be triggered during the L-H transition inside the otherwise safe $n_e$-$\delta_{n=1}$ operating space. In COMPASS, the corresponding EF threshold is observed at the overlap field magnitude of $\delta_{n=1} \cong 4 \cdot 10^{-4}$ and, importantly, does not appear to increase with density. Over the entire range of densities achieved, the EF penetration during the L-H transition is triggered at this moderate EF magnitude, although in steady-state plasma regimes it would only occur at densities as low as $2 \cdot 10^{19}$ m$^{-3}$ ($n_e / n_{GW} \sim 0.08$), as can be deduced from Figure 11.

Based on these results, it is recommended to separately consider the EF correction schemes also for such transient phases, especially in tokamaks that cannot afford a large number of disruptions (e.g. ITER). Moreover, due to the unfavorable size scaling [8], relatively low EF penetration thresholds are predicted for large tokamaks. Should this aspect of low EF threshold translate to the L-H transition, the H-mode accessibility may become limited for these devices.

To the best of our knowledge, no other experiment has quantified EF sensitivity during the L-H transition or other short transients so far. It is thus an important issue for future research to confirm if the reported disruptive conditions can be reproduced elsewhere, especially in larger contemporary devices. Initial attempts have been made to reproduce the preliminary COMPASS results [15], but these were focused on the role of the non-resonant EF during the L-H transition. In an L-H transition scenario at marginal NBI power in DIII-D [16], a shot with a non-resonant EF applied was disrupted at the time of the L-H transition in the reference shot, which was attributed to the effects of the NTV torque by the EF [37]. To date, there is no explicit relationship between the ITER EF threshold [8] and rotation, and as our paper shows, it is primarily the 2/1 mode rotation effect that leads to disruptions during the L-H transition.

The LM sensitivity to the rotation transient observed in COMPASS implies that the global plasma parameter scalings may also perform poorly in predicting the EF penetration thresholds during any other transient changes that affect the rotation dynamics. Assuming that a small-amplitude 2/1 tearing mode rotates with the electron fluid, it will evolve through any changes in the profiles of the bulk plasma rotation and the electron diamagnetic rotation. Both can be affected by auxiliary heating as well as by spontaneous changes in the transport processes.

Previously observed LMs induced by changing the level of auxiliary heating, especially the NBI with its significant torque, are likely caused by the mechanism reported in this work. In a DIII-D experiment with the net neutral beam torque injected in the counter-$I_p$ direction, special measures had to be taken to avoid LM formation during the rotation reversal [18]. In this case, the plasma equilibrium was similar to the ITER baseline, except for a higher $q_{95} \sim 4.3$, and the intrinsic plasma rotation was obviously in the co-$I_p$ direction. Note the difference with the case at $q_{95} \sim 3$ and $n_e \sim 4 \cdot 10^{19}$ m$^{-3}$ in COMPASS, where the core-deposited NBI did not significantly affect the 2/1 mode rotation (see Figure 10).

The natural variations of the intrinsic plasma rotation may be of greater concern for large tokamaks, given their relatively small external torque compared to the plasma inertia and the poor predictability of the intrinsic rotation behavior. Spontaneous rotation reversals in Ohmic L-mode, e.g. during density ramps, have been widely observed and often associated with the transition between linear (LOC) and saturated (SOC) Ohmic confinement ([38] and references therein). These intrinsic plasma rotation reversals have typically been confined to the core plasma. However, if the edge plasma rotation at least changes its magnitude, it can already reverse the 2/1 mode rotation due to the significant offset caused by the electron diamagnetic rotation in the edge.

Finally, one might expect a different behavior at higher $q_{95}$ than presented here. On the one hand, the sensitivity during the change of the edge plasma confinement may be limited to the case of low $q_{95}$, where the 2/1 mode rotation is more affected by the edge plasma dynamics. On the other hand, at higher $q_{95}$, the 2/1 mode rotation may be strongly affected by the transients in the core plasma, such as the formation of the internal transport barrier or the aforementioned reversal of the intrinsic plasma rotation. Therefore, special EF correction schemes for transient phases are potentially important at all values of $q_{95}$.

## 6. Conclusion

This work reports the first analysis of the disruptive locked mode instability triggered by the dynamics of the L-H transition. The sensitivity to $n = 1$ EF was investigated during both NBI-assisted and Ohmic L-H transitions in the COMPASS tokamak, at low $q_{95} \sim 3$ and $\beta_N \sim 1$, and using the EF coils located at the HFS of the vessel. Disruptive LM is regularly observed after the L-H transition over a wide range of densities up to $8 \cdot 10^{19}$ m$^{-3}$ ($n_e / n_{GW} > 0.2$), which is four times the critical density for LM in steady-state Ohmic plasma, as shown in Figure 11. A clear EF penetration threshold is quantified in terms of the total resonant overlap field $\delta_{n=1}$ [8] calculated by GPEC, at an amplitude of about $\delta_{n=1} \sim 4 \cdot 10^{-4}$. Notwithstanding the relatively limited dataset, the threshold shows no significant trend with density or applied torque, being predominantly a consequence of changes in the intrinsic rotation of the 2/1 mode during the transient phase of L-H. These results raise concerns for large tokamaks such as ITER, where the number of large disruptions is significantly limited, and urge against using any parametric EF penetration scaling derived from steady-state experiments to define the error field correction strategy in the entire discharge.

The aforementioned low threshold for EF penetration after the L-H transition in COMPASS is explained by the observed rapid decrease in the natural rotation frequency of the 2/1 mode during the first few ms of the H-mode, eventually reversing its direction from counter-$I_p$ to co-$I_p$ (Figure 9). This finding is in qualitative agreement with the rotation scaling expected from theory as well as several dedicated experiments. However, the assumption that the natural mode rotation can be predicted by global plasma parameters is challenged, as it clearly does not hold during the L-H transition in COMPASS, and presumably during other transient changes that affect the rotation, such as rotation reversals due to external or intrinsic torque.

The disruptive LM after the L-H transition is not only observed in 100% of the cases above the given EF penetration threshold, but is also identified after about 30% of the Ohmic L-H transitions below this threshold. The primary cause is proposed to be the resonant seeding of the 2/1 mode by the sawtooth precursor, which is allowed by the absence of rotation shear between the two modes after the Ohmic L-H transition in COMPASS. Consequently, only the NBI-assisted L-H transitions with no EF applied are found to be safe from disruptions.

This work highlights the importance of investigating the EF threshold also during the L-H transition and other transient changes, expanding on the existing experiments conducted in steady-state Ohmic or H-mode conditions. Our plan is to investigate whether the presented results can be replicated in other tokamaks and with different compositions of the applied EF. If the observed phenomenon proves to be robust, it will provide a sound guideline for optimizing the error field correction for the H-mode access in future devices such as ITER.

## Disclaimer

The views and opinions expressed herein do not necessarily reflect those of the ITER Organization.

## Acknowledgements

This work was supported by the Ministry of Education, Youth and Sports of the Czech Republic through project LM2023045, and by the U.S. Department of Energy, Office of Science, Office of Fusion Energy Sciences under Awards DE-SC0022270 and DE-SC0021968.